\pdfoutput=1
\documentclass{article}

\usepackage{preamble_arxiv}
\usepackage[accepted]{icml2020_arxiv}

\icmltitlerunning{Lorentz Group Equivariant Neural Network for Particle Physics}

\begin{document}

\twocolumn[
    \icmltitle{Lorentz Group Equivariant Neural Network for Particle Physics}
    
    
    
    
    \begin{icmlauthorlist}
        \icmlauthor{Alexander Bogatskiy}{b}
        \icmlauthor{Brandon Anderson}{a,f}
        \icmlauthor{Jan T.~Offermann}{b}
        \icmlauthor{Marwah Roussi}{b}
        \icmlauthor{David W.~Miller}{b,e}
        \icmlauthor{Risi Kondor}{a,c,d}
    \end{icmlauthorlist}
    \icmlaffiliation{a}{Department of Computer Science, University of Chicago, Chicago, IL, U.S.A.}
    \icmlaffiliation{b}{Department of Physics, University of Chicago, Chicago, IL, U.S.A.}
    \icmlaffiliation{c}{Department of Statistics, University of Chicago, Chicago, IL, U.S.A.}
    \icmlaffiliation{d}{Flatiron Institute, Simons Foundation, New York, NY, U.S.A.}
    \icmlaffiliation{e}{Enrico Fermi Institute, Chicago, IL, U.S.A.}
    \icmlaffiliation{f}{Atomwise, San Francisco, CA, U.S.A.} 
    
    \icmlcorrespondingauthor{Alexander Bogatskiy}{bogatsky@uchicago.edu}
    
    \icmlkeywords{Machine Learning, ICML, Physics, Equivariance, Particle Physics, Lorentz Group}
    
    \vskip 0.3in
]



\printAffiliationsAndNotice{}  

\begin{abstract}
We present a neural network architecture that is fully equivariant with respect to transformations under the Lorentz group, a fundamental symmetry of space and time in physics. The architecture is based on the theory of the finite-dimensional representations of the Lorentz group and the equivariant nonlinearity involves the tensor product. For classification tasks in particle physics, we demonstrate that such an equivariant architecture leads to drastically simpler models that have relatively few learnable parameters and are much more physically interpretable than leading approaches that use CNNs and point cloud approaches. The competitive performance of the network is demonstrated on a public classification dataset \cite{KasPleThRu19} for tagging top quark decays given energy-momenta of jet constituents produced in proton-proton collisions.
\end{abstract}

\section{Introduction}
\label{Introduction}

\looseness=-1
The success of CNNs as a method of computer vision has made clear the benefits of explicitly translationally equivariant neural network architectures: there are far fewer learnable parameters and these parameters are organized into much more interpretable structures. The ability to interpret convolutional kernels as images boosted our understanding of why and how such networks operate \cite{ZeilerFergus2014}.

However, there are many relevant problems that exhibit much more complex symmetries than flat images. Such problems may require or benefit from latent space representations that are intimately connected with the theory of the specific underlying symmetry group. Indeed, these symmetries are manifest in the data itself, as each data point is generated by a symmetric process or model. Following this approach, elegant architectures can be advised based on fundamental principles, and the ``building blocks'' of such architectures are greatly restricted by the imposed symmetries. This is a highly sought-after property in neural network design since it may improve generality, interpretability, and uncertainty quantification, while simplifying the model.

These general ideas have already led to the development of multiple equivariant architectures for sets (permutation invariance) \cite{ZaKoRaPoSS17}, graphs (graph isomorphisms), 3D data (spatial rotations) \cite{Monti2017Geometric}, and homogeneous spaces of Lie groups such as the two-dimensional sphere \cite{CohenSphericalICLR2018}. For more discussion and references see Section \ref{related work}.

Symmetries play a central role in any area of physics \cite{Frankel04}, and as such physics provides the widest variety of symmetry groups relevant in computational problems. In particular, high energy and particle physics involve symmetry groups ranging from $\mathrm{U}(1)$, $\SU(2)$ and $\SU(3)$ to the Lorentz group $\SO(1,3)$, and even more exotic ones like $\mathrm{E}_8$. Architectures that respect these symmetries can provide more sensible and tractable models, whose parameters may be directly interpreted in the context of known physical models, as in the case of CNNs.

Harmonic analysis provides two parallel but theoretically equivalent implementations of group equivariance in neural networks. The first is a natural generalization of CNNs to arbitrary Lie groups and their homogeneous spaces \cite{CohenWelli16}, where activations are functions on the group, the nonlinearity is applied point-wise, and the convolution is an integral over the group. The second approach works entirely in the Fourier space \cite{ThSmKeYLKR18,AndeHyKon19}, that is, on the set of irreducible representations of the group. It is the latter approach that we adopt in this work due to its direct applicability to vector inputs.

These approaches are general, but here we present the first specific application of a group equivariant architecture in physics. We focus on a particle physics application where the data typically contain the energy-momentum 4-vectors of particles produced in collision events at high energy particle accelerators such as the Large Hadron Collider (LHC) at CERN in Geneva, Switzerland, or by simulation software used to model the collision events. Probing elementary particle collisions at high energies is one of the best approaches to discover new small-scale fundamental phenomena, such as the discovery of Higgs boson at the LHC in 2012 \cite{Aad:2012tfa,Chatrchyan:2012xdj}. There the collisions occur 40 million times per second (40 MHz) between clouds of protons traveling at nearly the speed of light. Within each proton-proton bunch collision an average of $\mathcal{O}(30)$ individual pairs of protons collide and produce sprays of outgoing particles that are measured by complex detection systems. These \textit{detectors} -- such as the general-purpose ATLAS \cite{ATLAS2008} and CMS \cite{CMS2008} detectors -- have $\mathcal{O}(100M)$ individual sensors that record combinations of positions, trajectories, momenta, and energies of outgoing particles. The data obtained from these detectors must therefore be both filtered and processed by automated on-line systems.

The energy-momentum vector of a particle depends on the inertial frame of the observer, and the transformations between these frames are described by the Lorentz group $\mathrm{O}(1,3)$.  In addition to regular spatial rotations it contains the so-called \textit{Lorentz boosts}, which make the Lorentz group non-compact. The architecture presented below avoids computational difficulties associated with the non-compactness of the group by working entirely within its finite-dimensional representations. This choice is not only computationally efficient, but also physically sensible.


\section{Related Work}\label{related work}

\looseness=-1
There is a large body of work on equivariance in machine learning. Here we mention a few notable publications most closely related to the methods of our work. Equivariance in neural networks was first examined in applications involving finite groups, such as graph isomorphisms \cite{BruZarSzLe14, HenaBruLeC15} and permutations \cite{ZaKoRaPoSS17}. 
A general approach to group-convolutional neural networks was proposed in \cite{CohenWelli16}. Equivariant networks with respect to spacial translations and rotations were developed in \cite{WorGarTuBr16}.
For rotational symmetries of the 2-dimensional sphere, the importance of the Fourier space spanned by spherical harmonics was realized in  \cite{CohenSphericalICLR2018,EstAllMaDa18}. In \cite{WeGeWeBoCo18} this approach was extended to the entire Euclidean group $\mathrm{SE}(3)$.
A complete description of equivariant networks for scalar fields on homogeneous spaces of compact Lie groups was given in \cite{KondoTrive18}. It was later generalized to general gauge fields in \cite{CohWeiKiWe19, CoheGeiWei19}. 

The parallel approach, where even the nonlinear operations are performed equivariantly in the Fourier space of $\SO(3)$, was independently proposed in \cite{ThSmKeYLKR18} and \cite{Kondor18}.  Successful applications of these ideas in computer vision and chemistry were demonstrated in \cite{KondLinTri18, AndeHyKon19}. While the use of Lorentz-invariant quantities and Lorentz transformations in networks has been demonstrated in \cite{Butter:2017cot, Erdmann:2018shi}, our work provides the first equivariant neural network architecture for fundamental physics applications.

\section{Theory of the Lorentz group}\label{theory}

\paragraph{Lorentz transformations}

Particles moving in laboratories at velocities approaching the speed of light are described by the theory of special relativity. Its mathematical formulation is based on the postulate that space and time are unified into the 4-dimensional \textit{spacetime}, and the Euclidean dot product of vectors is replaced by the Minkowski, or Lorentzian, metric. In the standard Cartesian basis, this metric has the diagonal form $\mathrm{diag} (1,-1,-1,-1)$:
\[
(t,x,y,z)\cdot (t',x',y',z') = tt'\!-\!xx'\!-\!yy'\!-\!zz'=  \eta_{\mu\nu}x^\mu x'^\nu
\]
(here we set the speed of light equal to one and use the Einstein summation convention for repeated indices). Similarly, the energy and momentum of a particle are combined into the energy-momentum 4-vector whose square is also the mass squared of the particle:
\[
(E, p_x,p_y,p_z)^2 = E^2- p_x^2-p_y^2-p_z^2 = m^2.
\]
An \textit{inertial frame} in this spacetime is a choice of an orthonormal basis $\{\boldsymbol{e}_0,\boldsymbol{e}_1,\boldsymbol{e}_2,\boldsymbol{e}_3\}$, i.e.~\(\boldsymbol{e}_a\cdot \boldsymbol{e}_b = \eta_{ab}\), \(a,b=0,\ldots,3\). The components of the metric are the same in any such frame. The \textit{Lorentz group} is defined as the group of linear isomorphisms $\Lambda^\mu_\nu$ of the spacetime that map inertial frames to inertial frames, or equivalently, preserve the metric:
\[
\Lambda_\mu ^\lambda \eta_{\lambda\rho}\Lambda^\rho_\nu = \eta_{\mu\nu}.
\]
This group is denoted by $\mathrm{O}(1,3)$ and consists of 4 connected components distinguished by orientations of space and time. 

 
Often one further requires inertial frames to be positively oriented and positively time-oriented. That is, all orthonormal bases are similarly oriented and the timelike basis vector in each of them ($\boldsymbol{e}_0$) belongs to the future light cone (i.e.~its temporal component is positive). Restricting Lorentz transformations to only such frames (which amounts to requiring $\det \Lambda = 1$ and $\Lambda^0_0>0$), one obtains the \textit{proper orthochronous Lorentz group} $\SO^+(1,3)$, which is the connected component of the identity in $\mathrm{O}(1,3)$. From here on in this text, this is the group we will call the ``Lorentz group''. The basic principle of special relativity is that all laws of physics appear equivalent to observers in all inertial frames. This makes the Lorentz group the fundamental symmetry in relativistic physics.

The group $\mathrm{SO}(3)$ of spatial rotations (acting on $x,y,z$ in a chosen inertial frame) is a subgroup of the Lorentz group. In addition to these rotations, it contains the so-called \textit{Lorentz boosts} which transform between inertial frames of observers moving relative to each other at a relative velocity $\boldsymbol{\beta}=\boldsymbol{v}/c$ (in units of the speed of light $c$). Namely, given two inertial frames $\{\boldsymbol{e}_i\}_{i=0}^3$ and  $\{\boldsymbol{e}_i'\}_{i=0}^3$, the relative velocity vector $\boldsymbol{\beta}$ and the \textit{boost factor} $\gamma$ are defined by 
\(\boldsymbol{e}_0'=\gamma \boldsymbol{e}_0 + \sum_{i=1}^3 \gamma \beta_i\boldsymbol{e}_i.\)
Since $\boldsymbol{e}_0'$ has unit norm, the boost factor is related to $\boldsymbol{\beta}$ by $\gamma = (1-\beta^2)^{-1/2}$. Now, if one rotates the spatial axes so that $\boldsymbol{\beta}=(\beta,0,0)$ then the Lorentz transformation between these two frames is the matrix
\[\begin{pmatrix}
\gamma & -\gamma\beta & 0&0\\
-\gamma\beta & \gamma &0&0\\
0&0&1&0\\
0&0&0&1
\end{pmatrix}.\]
%
In the limit of speeds much lower than the speed of light, $\beta\to 0$,  $\gamma\to 1$, and this matrix becomes the identity matrix, returning us to Galilean mechanics. Therefore the appearance of the boost factor $\gamma$ is a signature of relativistic physics. Lorentz boosts are sometimes called ``hyperbolic rotations'' because their components can be expressed as $\gamma=\cosh\alpha$ and $\gamma\beta=\sinh\alpha$ in terms of a \textit{rapidity} $\alpha$. However, note that Lorentz boosts with more than one spatial dimension do not form a subgroup.

\paragraph{Representations}

Recall that a finite-dimensional representation of a Lie group $G$ is a finite-dimensional vector space $V$ with an action of the group via invertible matrices, that is, a smooth homomorphism $\rho :G\to \GL(V)$ (for some introductions to representations see \cite{BarutRac77, Hall15}, or \cite{Diaconis88} for a more applied focus). All activations in our neural network will belong to various finite-dimensional representations of the Lorentz group. Importantly, such representations are \textit{completely reducible}, which means that they are isomorphic to direct sums of irreducible representations (the isomorphism may not be unique). An irreducible representation (``irrep'') is one without any invariant subspaces, i.e.~subspaces $W\subset V$ such that $\rho(g)\cdot w\in W$ for all $g\in G$ and $w\in W$.
Writing the decomposition of $V$ as $V\cong \bigoplus_\alpha R_\alpha^{\oplus\tau_\alpha}$, where $R_\alpha$'s for different $\alpha$ are non-isomorphic irreps of $G$, we call $\tau_\alpha$ the \emph{multiplicity} of $R_\alpha$ in $V$. Written in terms of the subspaces $V_\alpha \cong R_\alpha^{\oplus \tau_\alpha}$, this decomposition $V=\bigoplus_\alpha V_\alpha$ is called the \emph{isotypic decomposition} of $V$.
Complete reducibility provides a natural basis for storing arbitrary representation vectors, therefore we will now review the classification of finite-dimensional irreps of $\SO^+(1,3)$.

The representation theory of the Lorentz group becomes slightly simpler if we pass to its universal covering group. For $\SO(3)$ the universal covering group is $\SU(2)$, also known as the 3-dimensional spin group, and for $\SO^+(1,3)$, which is isomorphic to the projective special linear group $\PSL(2,\bbC)$, it is $\SL(2,\bbC)$. Both of these are double covers, i.e.~we have $\SO(3)\cong \SU(2)/\{\pm I\}$ and $\SO^+(1,3)\cong \SL(2,\bbC)/\{\pm I\}$. Each irrep of the original group can be extended to an irrep of its double cover, but the double cover generally has more irreps \cite{GelMinSha63}. 
In physics, the extra ``double-valued'' irreps obtained by passing to the double cover are called \emph{spinor representations}.
Since $\SL(2,\bbC)$ is the complex form of $\SU(2)$, the finite-dimensional representations of these two groups are very closely related. These labels of the irreps are also known as \emph{highest weights} in representation theory. The irreps of $\SU(2)$ are indexed by the half-integer $l\in \bbN/2$ known as \textit{spin} in physics. We will denote these $(2l+1)$-dimensional irreps by $R^l$. Only the integer-spin $R^l$'s descend to irreps of $\SO(3)$.

\looseness=-1
The finite-dimensional irreps of the Lorentz group, or more generally the real irreps of its double cover $\SL(2,\bbC)$, are up to isomorphisms exactly the tensor products of representations of $\SU(2)$:
\[T^{(k,n)}=T^{(k,0)}\otimes T^{(0,n)}\coloneqq R^{k/2}\otimes \wb{R}^{n/2},\]
where $k,n$ are non-negative integers and the bar over $R^{n/2}$ indicates that this factor is acted upon by $\SL(2,\bbC)$ via the conjugated representation (explicitly shown below). The dimensions of these spaces are $\dim T^{(k,n)}=(k+1)(n+1)$. The irreps of the Lorentz group are those $T^{(k,n)}$ for which $k+n$ is even.

Recall that the action of $\SU(2)$ on its spin $l$ irrep is realized by the Wigner D-matrices $D^l(g)$, $g\in \SU(2)$. Due to the relation between $\SL(2,\bbC)$ and $\SU(2)$, it is easy to parametrize the group elements using Euler angles. Introduce
\begin{gather*}
    \alpha =\varphi+i \kappa ,\; \beta=\theta+i \epsilon,\; \gamma=\psi + i \varkappa,\\
    \varphi\in [0,2\pi),\; \theta\in [0,\pi],\; \psi\in [0,2\pi),\;  \kappa,\epsilon,\varkappa\in\bbR
\end{gather*} 
($\beta$ and $\gamma$ should not be confused with the velocity and boost factors from special relativity). These variables provide non-degenerate coordinates on $\SL(2,\bbC)$, identifying it with the space $S^3\times \bbR^3$. Any unimodular matrix $a\in \SL(2,\bbC)$ can be factorized as
\begin{multline*}
a(\alpha,\beta,\gamma)=\\ 
\begin{pmatrix}e^{i\alpha/2} & 0\\
0 & e^{-i\alpha/2}
\end{pmatrix}\begin{pmatrix}\cos\frac{\beta}{2} & i\sin\frac{\beta}{2}\\
i\sin\frac{\beta}{2} & \cos\frac{\beta}{2}
\end{pmatrix}\begin{pmatrix}e^{i\gamma/2} & 0\\
0 & e^{-i\gamma/2}
\end{pmatrix}\,,\label{Euler factorization}
\end{multline*}
which is the complexification of the Euler factorization. Real angles parametrize the $\SU(2)$ subgroup, whereas the imaginary parts are essentially the rapidities parametrizing Lorentz boosts. This formula also expresses the so called \emph{fundamental}, or \emph{defining}, representation of $\SL(2,\bbC)$ acting on $T^{(1,0)}\cong \bbC^2$.

Furthermore, it is clear that the action of $\SL(2,\bbC)$ on the irrep $T^{(k,0)}$ is given exactly by the analytically continued Wigner D-matrix of $\SU(2)$ spin $k/2$. Similarly, the action on $T^{(0,n)}$ is given by the conjugate representation of spin $n/2$. The conjugate representation of $\SL(2,\bbC)$ of spin $1/2$ (the conjugate fundamental one) is given by $a\mapsto \epsilon\,\overline{a}\,\epsilon^{-1}$ where $\epsilon$ is the 2D Levi-Civita tensor. It is easy to check that 
\[\epsilon\,\overline{a(\alpha,\beta,\gamma)}\,\epsilon^{-1}=\overline{a(-\alpha,\beta,-\gamma)}\]
(here the bar denotes complex conjugation). Combining these, we see that the action on $T^{(k,n)}$ corresponds to the tensor product of two Wigner D-matrices:
\[D^{k/2}(\alpha,\beta,\gamma)\otimes \overline{D^{n/2}(-\alpha,\beta,-\gamma)}.\]
For instance, on the fundamental representation of $\SO^+(1,3)$, for which $k\!=\!n\!=\!1$, these matrices are exactly the $4\times 4$ Lorentz transformations (the defining representation of $\SO^+(1,3)$).

\section{Principles of Equivariant Networks}\label{principles}

\paragraph{Equivariant Universal Approximation}\label{EUA}

Given two representations $(V,\rho)$ and $(V',\rho')$ of a group $G$, a map $F:V\to V'$ is called equivariant if it \textit{intertwines} the two representations, that is:
\[
F(\rho(g)\cdot v)=\rho'(g)\cdot F(v),\quad v\in V,\; g\in G.
\]
Our goal is to design an architecture that can learn arbitrary equivariant maps between finite-dimensional representations of the Lorentz group. Even though the application described below requires only invariant outputs, the general way to achieve this is with an internally equivariant structure. First and foremost, this means having activations that are elements of linear representations of the group.

It was shown in \cite{Yarotsky18} that an arbitrary equivariant map between two completely reducible representations can be approximated by linear combinations of copies of a non-polynomial function $\sigma$ applied to linear functions of $G$-invariants, with coefficients from a basis of $G$-equivariants (see Supplementary Material Section \ref{Sec. Universality theorem} for more details). Importantly, these polynomial invariants and equivariants are multiplicatively generated by a finite basis. This approximation theorem reduces our task to generating arbitrary polynomial invariants and equivariants for finite-dimensional representations of $\SL(2,\bbC)$.
In the Supplementary Material we show an extended version of the $G$-equivariant universal approximation theorem from \cite{Yarotsky18}, which we paraphrase in simple terms here.
%
\begin{thm}\label{Thm}
    Given two completely reducible finite-dimensional representations $V$ and $U$ of a Lie group $G$, which can be $\SU(2)$, $\SO(3)$, $\SL(2,\bbC)$, or $\SO^+(1,3)$, 
    any equivariant map $\wt{f}:V\to U$ (including invariant maps for which $U\cong \bbR$) can be approximated by a feed-forward neural network with vector activations belonging to finite-dimensional representations of $G$ that can iteratively perform the following operations:
    \begin{enumerate}
        \item Take tensor products of elements of representations of $G$;
        \item Decompose tensor representations into isotypic components using the Clebsch-Gordan decomposition;
        \item Apply equivariant linear maps between representations of $G$ (as detailed in Section \ref{linear maps}), including projections onto specific isotypic components;
        \item Apply arbitrary sub-networks (such as multilayer perceptrons) to any $G$-invariants appearing after any of the above operations.
    \end{enumerate}
\end{thm}

Note that this theorem is a ``Fourier space'' statement (i.e.~regarding networks based on irreps) extending the ``real-space'' characterization theorem proven in \cite{KondoTrive18}.

\paragraph{Equivariant Linear Maps}\label{linear maps}

Now that we have established that tensor products are sufficient as the equivariant nonlinearity, we need to specify the form of equivariant learnable linear operations. Given a completely reducible representation $V$ of $G$, we first find a linear isomorphism on $V$ that represents it as a direct sum of its isotypic components: $V=\bigoplus_\alpha V_\alpha$ (the sum is taken over the labels $\alpha$ of all finite-dimensional irreps). Typically for us this isomorphism will be given by a Clebsch-Gordan operator. Each component $V_\alpha$ is itself isomorphic to a direct sum of zero or more copies of an irrep $R_\alpha$: $V_\alpha= R_\alpha^{\oplus \tau_\alpha}$. We call $\tau_\alpha$ the multiplicity of the irrep $R_\alpha$ in $V$. Now suppose the target representation can be similarly decomposed as $U=\bigoplus_\alpha R_\alpha^{\oplus \tau_\alpha'} $. Then, as was proven in \cite{KondoTrive18} by an extension of Schur's Lemma, all linear equivariant maps $W: V\to U$ are parametrized by a collection of $\tau_\alpha'\times \tau_\alpha $ matrices 
\[W_\alpha\in \mathrm{Mat}(\tau_\alpha',\tau_\alpha),\]
each of which acts on the list of irreducible components within an isotypic component $V_\alpha$. This characterization (but not $W_\alpha$'s themselves) is independent of the choice of decompositions of the $V_\alpha$'s into irreducible components. 

As demonstrated in \cite{CohenWelli16}, the restriction to equivariant linear layers, compared to a general fully connected linear layer, leads to significantly fewer learnable parameters (depending on the representations at hand). Perhaps most importantly, \emph{the loss function itself is $G$-invariant}. This means that if we transform the training samples $(\boldsymbol{x}_i,\boldsymbol{y}_i)$ by group elements $g_i\in G$, the trained weights $W$ of an equivariant network will remain the same. In this sense, the weights are $G$-invariant, which makes them potentially interpretable as physical quantities.


\begin{figure*}
    \centering
    \includegraphics[scale=0.7]{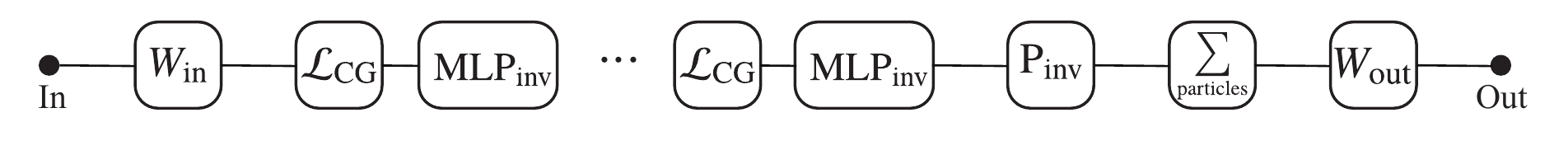}
    \caption{An elementary flow chart of LGN with Lorentz-invariant outputs. $W_{\mathrm{in}}$ is the linear input layer. It is followed by iterated CG layers $\mathcal{L}_\mathrm{CG}$ defined in (\ref{master formula}) alternated with perceptrons $\mathrm{MLP_{inv}}$ acting only on Lorentz invariants. The output layer projects onto invariants using $\mathrm{P_{inv}}$, sums over particles for permutation invariance, and applies a linear layer. $W_{\mathrm{in}}$, $\mathrm{MLP_{inv}}$ and  $\mathrm{P_{inv}}$ act on each particle separately, but have the same values of parameters across all particles.}
    \label{fig:diagram}
\end{figure*}

\paragraph{Particle Interactions.}\label{particle interactions}

As an elementary example of learnable Lorentz-invariant quantities in particle physics, the electron-muon scattering matrix element for  initial and final 4-momenta $p_1, p_3$ of the electron, and  initial and final 4-momenta $p_2, p_4$ of the muon, is given by
\[\mathcal{M}^2\propto \frac{g_c^4\left[p_1\cdot p_3+m_e^2\right]\left[p_2\cdot p_4+m_\mu^2\right]}{((p_1-p_3)^2-m_\gamma^2)^2}.\]
Here the dot products are taken with respect to the Minkowski metric, $m_e^2=p_1^2$ and $m_\mu^2=p_2^2$ are the masses, and $g_c$ is an interaction strength parameter. Dot products are the invariant parts in the isotypic decompositions of tensor products of two 4-vectors, therefore a quantity of this kind can be very efficiently learned by an equivariant network if physically appropriate nonlinear activation functions are chosen. More complicated processes would involve higher nonlinearities like $(p_1^\mu p_2^\nu - p_1\cdot p_2 \eta^{\mu\nu})^2$, which require several tensor products to be generated. 

When a particle decay event produces hundreds of observed particles, generating all relevant Lorentz invariants (and even more so equivariants) up to a fixed polynomial degree quickly becomes an intimidating task that begs for a procedural solution. This is exactly the goal of our architecture.

\section{Clebsch-Gordan product}

\looseness=-1
The main nonlinearity in our equivariant architecture is the tensor product followed by a decomposition into irreducibles. This decomposition is known as the Clebsch-Gordan (CG) decomposition, and its coefficients in a certain canonical basis are called CG coefficients. 
We introduce the notation for the coefficients and a final formula for the CG coefficients of the Lorentz group here, but leave the details and derivations to the Supplementary Material. A reference for this material as regards $\SU(2)$ and the Lorentz group is \cite{GelMinSha63}.

\paragraph{Rotation group}

Let $R_{l_1}$ and $R_{l_2}$ be irreps of $\SU(2)$ of half-integer weights (spins) $l_1$ and $l_2$, respectively. Their product $R_{l_1}\otimes R_{l_2}$ decomposes via an isomorphism into a direct sum $\bigoplus_l\wt{R}_l$, where $\wt{R}_l$ are also copies of irreps of $\SU(2)$ and $l$ ranges from $|l_1-l_2|$ to $l_1+l_2$ with unit step. This isomorphism is called the Clebsch-Gordan map
\[B: \quad \bigoplus_{l=|l_1-l_2|}^{l_1+l_2}\wt{R}_l \to  R_{l_1}\otimes R_{l_2}.\]
Since $\SU(2)$ is compact, its finite-dimensional representations can be assumed to be unitary with respect to the Euclidean norms on $\bbC^n$ (the resulting representation matrices are called Wigner D-matrices), therefore we can always choose $B$ so that it is orthogonal.

For an arbitrary representation of $\SU(2)$ we define the \emph{canonical basis} in it by $e_{l,m}$ where $l$ ranges over the weights of the irreps contained in the representation, and for each $l$, the index $m$ ranges over $-l,-l+1,\ldots,l$. Therefore the product space $R_{l_1}\otimes R_{l_2}$ has a basis induced from the respective canonical bases of the factors,
\[e_{l_1,m_1}\otimes e_{l_2,m_2},\quad m_1=-l_1,\ldots,l_1,\quad m_2=-l_2,\ldots,l_2,\]
and the space $\bigoplus_l\wt{R}_l$ naturally has the canonical basis
\[\wt{e}_{l,m},\quad l=|l_1-l_2|,\ldots,l_1+l_2,\quad m=-l,\ldots,l.\]
The CG coefficients $B^{l_1,m_1;l_2,m_2}_{l,m}$ are defined as the components of the CG map in these two bases:
\[B:\quad \wt{e}_{l,m}\mapsto \sum_{m_1,m_2} B_{l,m}^{l_1,m_1;l_2,m_2}e_{l_1,m_1}\otimes e_{l_2,m_2}.\label{def of CG coeff}\]
The summation is taken over all free indices occurring twice (and we will often omit mentioning them) over the ranges $|m_1|\leq l_1$, $|m_2|\leq l_2$,
however $B_{l,m}^{l_1,m_1;l_2,m_2}$ vanishes whenever $m_1+m_2\neq m$ (see e.g.~\citep[Ch. 4]{VilenKlimy95} for more on representation theory and CG coefficients of some classical groups).

\paragraph{Lorentz group}

The proper orthochronous Lorentz group $\SO(1,3)^+$ is isomorphic to the projective special complex linear group $\PSL(2,\bbC)$.
The Clebsch-Gordan map in this case is the isomorphism
\[H:\quad  \bigoplus_{k,n}\wt{T}^{(k,n)} \to T^{(k_1,n_1)}\otimes T^{(k_2,n_2)},\]
where the sum on the left is over
\begin{gather*}
    k=|k_1-k_2|,|k_1-k_2|+2,\ldots,k_1+k_2,\\ n=|n_1-n_2|,|n_1-n_2|+2,\ldots,n_1+n_2.
\end{gather*}

When an irrep $T^{(k,n)}$ of $\SL(2,\bbC)$ is viewed as a representation of its subgroup $\SU(2)$, it decomposes into the direct sum of irreps (with unit multiplicities)
\(T^{(k,n)}\cong \bigoplus_{l=|k-n|/2}^{(k+n)/2} R_{l}.\)
This way, $T^{(k,n)}$ admits a \emph{canonical basis}
\[e^{(k,n)}_{l,m},\quad l=|k-n|/2,\ldots,(k+n)/2;\quad m=-l,\ldots,m.\]
In this basis, we define the CG coefficients for the Lorentz group by 
\[H:\; \wt{e}^{(k,n)}_{l,m}\mapsto 
\sum H_{(k,n),l,m}^{(k_1,n_1),l_1,m_1;(k_2,n_2),l_2,m_2}e^{(k_1,n_1)}_{l_1,m_1}\otimes e^{(k_2,n_2)}_{l_2,m_2}.\]

The CG coefficients can be expressed in terms of the well known coefficients for $\SU(2)$ introduced above:
\begin{multline}
    H_{(k,n),l,m}^{(k_1,n_1),l_1,m_1;(k_2,n_2),l_2,m_2}=
    \sum_{m_1',m_2'}\\ B^{\frac k2,m_1'+m_2';\frac n2,m-m_1'-m_2'}_{l,m}
    B^{\frac{k_1}{2},m_1';\frac{k_2}{2},m_2'}_{\frac{k}{2},m_1'+m_2'}
    B^{\frac{n_1}2,m_1-m_1';\frac{n_2}2,m_2-m_2'}_{\frac n2,m-m_1'-m_2'}\times \\
    \times B^{\frac{k_1}2,m_1';\frac{n_1}2,m_1-m_1'}_{l_1,m_1}
    B^{\frac{k_2}2,m_2';\frac{n_2}2,m_2-m_2'}_{l_2,m_2},
\end{multline}

where the sum is taken over the following range of indices:
\begin{gather*}
    -\frac k2 \leq m_1'+m_2' \leq \frac k2,\quad
    m-\frac n2\leq m_1'+m_2'\leq m+\frac n2,\nonumber\\
    |m_1'|\leq \frac{k_1}2,\quad m_1-\frac{n_1}2\leq m_1'\leq \frac{n_1}2+m_1,\\
    |m_2'|\leq \frac{k_2}2,\quad m_2-\frac{n_2}2\leq m_2'\leq \frac{n_2}2+m_2.\nonumber
\end{gather*}
As always, the CG coefficients vanish when $m_1+m_2\neq m$. We provide more details on the derivation and computational implementation of this important formula in the Supplementary Material.

\section{Equivariant Architecture (LGN)}\label{LGN section}

We now describe the specific architecture that we applied to the problem outlined in Section \ref{Experiment}. We call it the Lorentz Group Network (LGN).

\paragraph{Permutation Invariance}

Since the physics is independent of the labeling we put on the observed particles, the output of the network must also be invariant with respect to the permutations of the inputs. For our architecture this means that all learnable weights must be independent of the index of the input, and the simplest way to achieve it is with sums over that index at appropriate stages in the network. These sums are a key part of the architecture described here.

\paragraph{Input layer}

The inputs into the network are 4-momenta of $N_{\mathrm{obj}}$ particles from a collision event, and may include scalars associated with them (such as label, charge, spin, etc.). That is, the input is a set of vectors living in a $\left(T^{(0,0)}\right)^{\oplus \tau_0}\oplus T^{(1,1)}$ representation of the Lorentz group. Here, $\tau_0$ is the number of input scalars. In this case, $\tau_0=2$ and the corresponding scalars are the mass of the particle and a label distinguishing observed decay products from the collider beams.

The input layer is simply a fully-connected linear layer acting on the inputs and producing $N^{(0)}_{\mathrm{ch}}$ (number of ``channels'' at layer 0) vectors in each irreducible component. This layer acts on each input separately but the weights are shared between them to enforce permutation invariance. 

\paragraph{CG Layers}

At the end of the input layer, we have $N_{\mathrm{obj}}$ activations $\mathcal{F}^{(0)}_i,\; i=1,\ldots,N_{\mathrm{obj}},$ living in $\left(T^{(0,0)}\oplus T^{(1,1)}\right){}^{\oplus N^{(0)}_{\mathrm{ch}}}$. We then apply a CG layer, iterated $N_{\mathrm{CG}}$ times, that performs tensor products, Clebsch-Gordan decompositions, and a learnable linear operation.

Assume that at the start of the $p$-th CG layer (starting with $p=0$) we have $N_{\mathrm{obj}}$ activations $\mathcal{F}^{(p)}_i$ living in some representations of the Lorentz group (in fact our architecture guarantees that the representation is independent of $i$). The CG layer updates these activations $\mathcal{F}^{(p)}_i\mapsto \mathcal{F}^{(p+1)}_i$ according to the update rule
\begin{multline}
    \mathcal{F}^{(p+1)}_i =\mathcal{L}_\mathrm{CG}\left(\mathcal{F}^{(p)}\right)_i=W\cdot  \left(\mathcal{F}^{(p)}_i\oplus \mathrm{CG}\left[\mathcal{F}^{(p)}_i\right]^{\otimes 2} \oplus\right. \\
    \left.\oplus \mathrm{CG}\left[\sum_j f(p_{ij}^2) p_{ij}\otimes \mathcal{F}^{(p)}_j\right]\right). \label{master formula}
\end{multline}
The Clebsch-Gordan operator $\mathrm{CG}$ follows every tensor product, and we are able to keep only the first few isotypic components to control memory usage. The last term models two-particle interactions via the pair-wise differences $p_{ij}=p_i-p_j$ while ensuring permutation invariance. The scalar coefficients $f(p_{ij}^2)$ in this sum involve a function $f:\bbR\to \bbR$ with some learnable parameters, which weights the interactions of the $i$'th particle with other particles. The second term models a self-interaction of the $i$'th particle, and the first term simply stores the activation from the previous layer. $W$ (also independent of $i$ to ensure permutation invariance) is the equivariant learnable operator described earlier, and it mixes each isotypic component to a specified number $N_{\mathrm{ch}}^{(p+1)}$ of channels. This choice controls the size of resulting vectors without breaking permutation invariance or Lorentz equivariance. To minimize computations, tensor products are performed channel-wise, which doesn't affect expressive ability due to the presence of learnable linear operators mixing the channels.

\paragraph{MLP Layers}

Since Lorentz invariants can be freely transformed by arbitrary nonlinear functions without fear of breaking Lorentz symmetry, we can use traditional scalar neural networks each time any invariants are generated in our equivariant network. Namely, at the end of each CG layer we apply a multilayer perceptron to the $\left(T^{(0,0)}\right)^{\oplus N_{\mathrm{ch}}{}^{(p)}}$ isotypic component. It takes $N_{\mathrm{ch}}^{(p)}$ scalar inputs and produces the same number of outputs. The parameters of this perceptron are shared across all $N_{\mathrm{obj}}$ nodes in the CG layer. Adding these layers ensures that the layers of the network are non-polynomial.

\paragraph{Output Layer}

For permutation invariance, the output layer must take an arithmetic sum of the $N_{\mathrm{obj}}$ activations produced after the last CG layer. For a classification task, we are only interested in Lorentz-invariant outputs, therefore the output layer extracts the invariant isotypic component of this sum, and applies a final fully connected linear layer $W_{\mathrm{out}}$ to the $N_{\mathrm{ch}}^{(N_{\mathrm{CG}})}$ scalars, producing 2 scalar weights for binary classification:
\[\vec{w}_{out} = W_{\mathrm{out}}\cdot \left(\sum_i \mathcal{F}_i^{(N_{\mathrm{CG})}}\right)_{(0,0)},\]
where $\left(\right)_{(0,0)}$ denotes a projection onto the spin-0 isotypic component (i.e.~Lorentz invariants).

\section{Experiments}\label{Experiment}

We have tested the covariant LGN architecture on the problem of \textit{top tagging}. This is a classification task that aims to identify top quark ``jets'' among a background of lighter quarks. Since the classification task is independent of the inertial frame of the observer,  the outputs of the classifier should be Lorentz invariants.


\paragraph{Jets}

\begin{figure}[ht]
\begin{center}
\centerline{\includegraphics[width=0.4 \columnwidth]{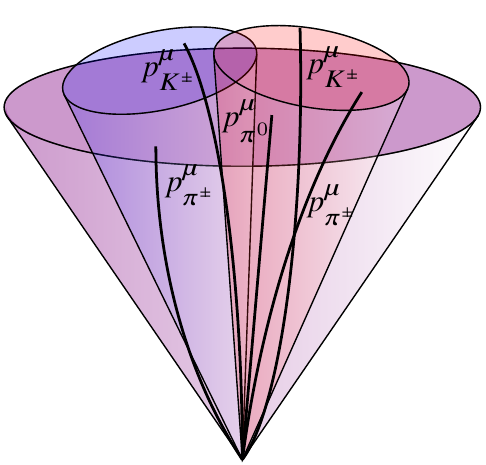}}
\caption{An example jet -- as shown, different jet parameters (such as radius) may result in different clustering.}
\label{jets figure}
\end{center}
\vskip -0.2in
\end{figure}

As explained in \citep{Salam:2009jx}, high energy quarks produced in particle collisions lose energy through a cascading gluon emission process -- a so-called \textit{parton shower} -- due to the structure of Quantum Chromodynamics (QCD), and eventually form stable hadrons that may be detected and measured. The lab frame in which those measurements are made may significantly differ from the parent quark's center-of-mass frame due to a Lorentz boost. In such a Lorentz-boosted lab frame, the parton shower process forms a collimated spray of energetic hadrons, depicted in \ref{jets figure}, known as a \textit{jet}. The jet 4-vector is related to that of its parent quark, as is the spatial and kinematic \textit{structure} of the particles contained within the jet. A crucial task in collider physics is discerning the species of quark that has given rise to a particular jet. Approaches to this task involve the use of theory-inspired analytic observables, feed-forward neural networks, CNNs, recurrent neural networks, point clouds, and more. For a recent and comprehensive review, see \cite{Larkoski:2017jix}.

\paragraph{Dataset}

We perform top tagging classification experiments using the LGN architecture and the publicly available reference dataset \cite{KasPleThRu19}. This dataset contains 1.2M training entries, 400k validation entries and 400k testing entries.
Each of these entries represents a single jet whose origin is either an energetic top quark, or a light quark or gluon. The events were produced with center-of-mass energy $\sqrt{s} = \SI{14}{\tera\electronvolt}$, using the PYTHIA Monte Carlo event generator \cite{Sjostrand:2014zea}. The ATLAS detector response was modeled with the DELPHES software package \cite{deFavereau:2013fsa}.

The jets in the reference dataset are clustered using the anti-$k_t$ algorithm \cite{Cacciari:2008gp}, with a radius of $R=1$, where $R$ is measured in  $(\eta,\phi)$ coordinates. For each jet, the energy-momentum 4-vectors are saved in Cartesian coordinates $(E,p_x,p_y,p_z)$ for up to $200$ constituent particles selected by the highest transverse momentum $p_T = \sqrt{p_x^2 + p_y^2}$, where the colliding particle beams are aligned along the $z$-axis. Each jet contains an average of 50 particles, and events with less than 200 are zero-padded.

The 4-momenta in the dataset are all scaled by a uniform factor at the input to the network to avoid overflows and losses. As an extra pre-processing step, we add the proton beams to the list of particles as two 4-momenta of the form $(2,0,0,\pm 1)\,\SI{}{\giga\electronvolt}$\footnote{Special thanks to Jesse Thaler for this suggestion.}. The purpose of this is to fix an axis in each sample event, thus establishing a symmetry-breaking relationship between different samples. The energy (chosen to be \SI{2}{\giga\electronvolt}) of these beams is somewhat arbitrary. Since these beams are distinct from the actual decay products in the dataset, we add a label to each particle, equal to $+1$ for the proton beams and $-1$ for all other particles. These labels are treated as Lorentz scalars.

\paragraph{Hyperparameters}

For training, we performed a manual grid search. The main parameters are the number of CG layers, the highest irrep kept after each tensor product, and the numbers of channels. For top tagging, we found it sufficient to keep $T^{(k,n)}$ with $k,n\leq 2$, which means that the highest irrep is the 9-dimensional $T^{(2,2)}$ and the remaining irreps are $T^{(0,0)}$, $T^{(2,0)}$, $T^{(0,2)}$, and $T^{(1,1)}$. There were 3 CG layers, and the numbers of channels were chosen as $N_{\mathrm{ch}}^{(0)}=2$, $N_{\mathrm{ch}}^{(1)}=3$, $N_{\mathrm{ch}}^{(2)}=4$, $N_{\mathrm{ch}}^{(3)}=3$. The internals of the network are based on complex arithmetic, so these numbers should be doubled to count the number of real components.

The MLP layer after the $p$-th CG layer had 3 hidden layers of width $2 N_{\mathrm{ch}}^{(p)}$ each and used the ``leaky ReLU'' activation function. 
The scalar function $f$ in \ref{master formula} was a learnable linear combination of 10 basis ``Lorentzian bell''-shaped curves $a+1/(1+ c^2x^2)$ with learnable parameters $a,b,c$ (each taking 10 values).
The input 4-momenta were scaled by a factor of $0.005$ to ensure that the mean values of the components of all activations would be order 1.

\looseness=-1
All weights were initialized from the standard Gaussian distribution. To ensure that activations stay of order one on average, the weights $W$ were scaled down by a factor $N_{\mathrm{ch}}^{(p)}/\tau_{(k,n)}$, where $\tau_{(k,n)}$ is the multiplicity of the $T^{(k,n)}$ irrep in the input to $W$. This ensures that $W$ does not distort the values of the activations in higher irreps by orders of magnitude, making the contributions of various irreps unbalanced.

\paragraph{Performance and Cost}

The architecture was coded up using PyTorch and trained on two clusters with GeForce RTX 2080 GPU's. Each training session used one GPU and with the hyperparameters listed above it used about 3700MB of GPU memory with a mini-batch size of 8 samples. The wallclock time was about 7.5 hours per epoch, and our models were trained for 53 epochs each.

We compare the performance of our network to some of the other competitors (for a review see \cite{KasiePlehn19}). For each of these binary classifiers, we report four characteristics: the accuracy, the Area Under the Curve (AUC) score, the background rejection $1/\epsilon_B$ at the signal efficiency of $\epsilon_S=0.3$ ($\epsilon_B$, $\epsilon_S$ are also known as the false positive and the true positive rates, respectively), and the number of trainable parameters. Higher accuracy, AUC and $1/\epsilon_B$ are considered better. The mean and standard deviation in these metrics for LGN are reported based on 4 independent trained instances of the model.

\begin{table}[h]
\vskip -0.2in
\caption{Performance comparison between LGN and other top taggers that were measured in \cite{KasiePlehn19}. Each performance metric is an average over an ensemble of networks, with the uncertainty given by the standard deviation.}
\label{sample-table}
\vskip 0.15in
\begin{center}
\begin{small}
\begin{tabular}{lcccr}
\toprule
Architecture & Accuracy & AUC & $1/\epsilon_B$ & \#Param \\
\midrule
ParticleNet   & 0.938 & 0.985 & 1298 $\pm$ 46 & 498k \\
P-CNN    & 0.930 & 0.980 & 732 $\pm$ 24 & 348k \\
ResNeXt    & 0.936 & 0.984 & 1122 $\pm$ 47 & 1.46M \\
EFP    & 0.932 & 0.980 & 384 & 1k \\
EFN    & 0.927 & 0.979 & 633 $\pm$ 31 & 82k \\
PFN  & 0.932 & 0.982 & 891 $\pm$ 18 & 82k \\
TopoDNN   & 0.916 & 0.972 & 295 $\pm$ 5  & 59k \\
\midrule 
LGN   & 0.929 & 0.964 & {435 $\pm$ 95}   & {4.5k} \\
      & $\pm$ .001 & $\pm$ 0.018 & & \\
\bottomrule
\end{tabular}
\end{small}
\end{center}
\vskip -0.2in
\end{table}
 
 The references for the algorithms listed here are: ParticleNet \cite{ParticleNet}, P-CNN \cite{PCNN2}, ResNeXt \cite{ResNeXt}, EFP \cite{EFP}, EFN and PFN \cite{EFN}, TopoDNN \cite{TopoDNN}. We should highlight EFP which constructs a special linear basis of polynomial observables that satisfy the so-called IRC-safety requirement in particle physics, and EFN which extends this idea to deep neural networks.
 
While our results do not match the state of the art, our model uses between $10-1000\times$ fewer parameters. More analysis of training and performance is provided in the Supplementary Material.

\begin{figure}[ht]
\begin{center}
\centerline{\includegraphics[width=\columnwidth]{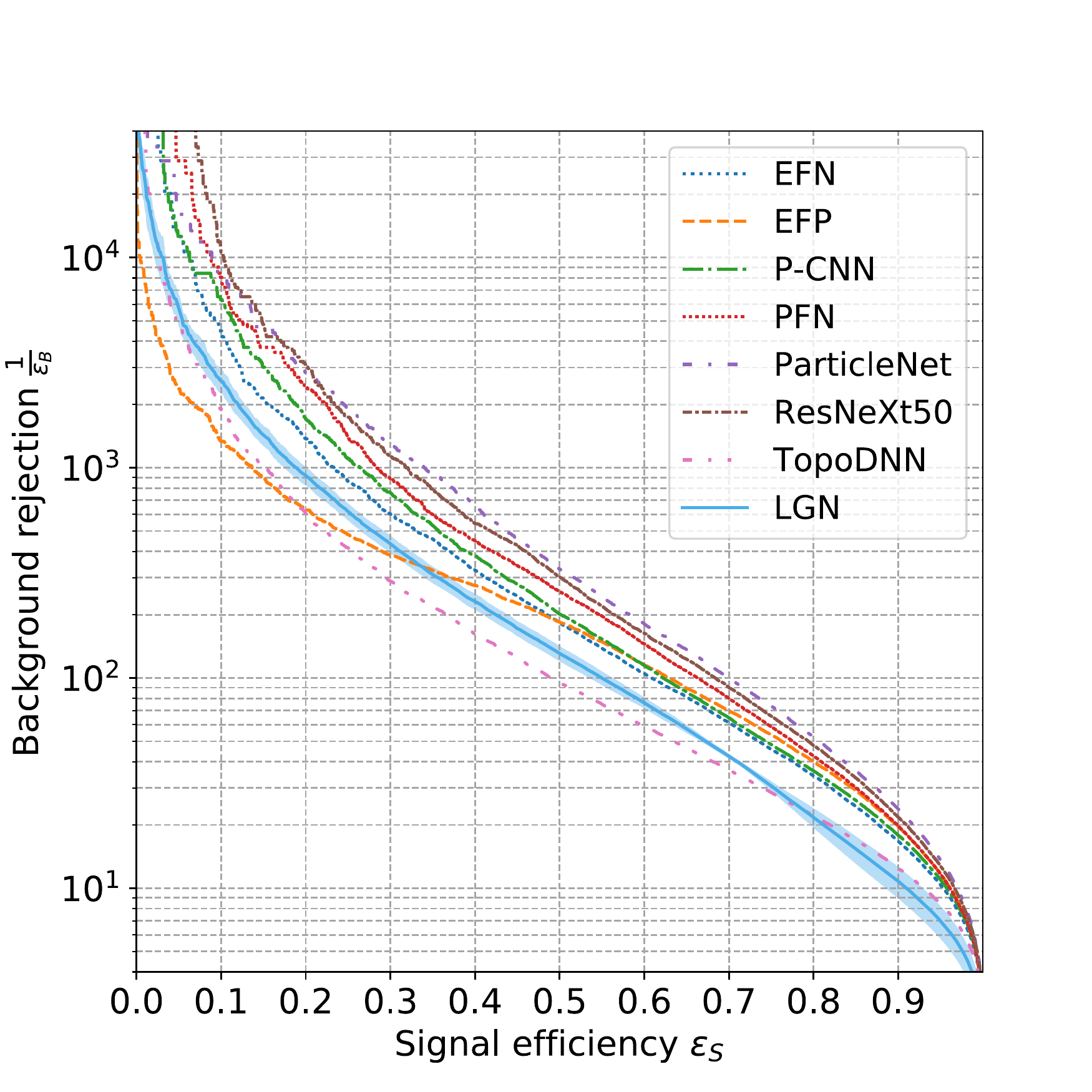}}
\caption{A comparison of an averaged ROC for LGN, against a sample of other top taggers. Higher is considered better. The ROC for LGN was sampled over the 4 trained instances of the model, with the error band width given by the standard deviation.}
\label{ROC Plot}
\end{center}
\vskip -0.2in
\end{figure}

\section{Conclusion}\label{conclusion}

We have developed and successfully applied a Lorentz-equivariant architecture for a classification task in particle physics, top tagging. We believe this is the first application of a fully Fourier space equivariant architecture in physics, following an chemistry application in \cite{AndeHyKon19}, and an important early step in building a family of physics-informed machine learning algorithms based on group theory. Symmetries have always been a central part of model-building in physics, and this work only further demonstrates the sharp need for symmetry- and geometry-based approaches to machine learning for scientific applications.

The performance of our neural network shines especially in terms of the number of learnable parameters. The trade-off is that an equivariant architecture takes more time to develop and its evaluation is more computationally intensive. However, once developed for a specific symmetry group, such as the Lorentz group or $\SL(2,\bbC)$ in our case, it is broadly applicable to many problems with the same symmetry at a very low development cost.

This network allows for many promising extensions in the context of particle physics. Future work will explore additional particle information such as charge and spin. The parameters of the model, which are Lorentz-invariant by construction, should be interpreted as physical quantities describing the particle decays. Permutation invariance can be further extended to permutation covariance. Another exciting problem is applying the network to regression tasks such as measuring masses of particles, or even 4-momenta. Finally, one could combine multiple symmetries such as the symmetry group of the Standard Model of physics (which includes $\mathrm{U}(1)$, $\SU(2)$ and $\SU(3)$).

\section{Acknowledgements}

We thank Jesse Thaler for helpful advice on the training process. We acknowledge the significant support from the University of Chicago’s Research Computing Center, Department of Computer Science, and especially the Center for Data and Computing (CDAC) for supporting this work through its Data Science Discovery Grant program. Finally, we thank the Flatiron Institute for hosting some of the authors during parts of the preparation of this paper. R.~Kondor was supported by DARPA grant number HR00111890038.


\setlength\bibhang{0pt} 
\setlength\bibitemsep{0pt} 
\setstretch{0.85} 
\setquotestyle{english}
\printbibliography[heading=bibintoc]
\setstretch{1} 

\appendix

\section{Clebsch-Gordan (CG) coefficients}

Here we provide further details on the Clebsch-Gordan decompositions for $\SU(2)$ and $\SL(2,\bbC)$ and their computer implementation. A good reference for this material is the book by \citet{GelMinSha63}, however that book contains some errors that lead to an incorrect expression for the CG coefficients of the Lorentz group. Since we are not aware of a reference with the correct formulas, we re-derive them here.

We first make a note about the inverse CG mapping for $\SU(2)$.
By orthogonality of the CG mapping, we have its inverse
\[B^{-1}=B^T:\quad  R_{l_1}\otimes R_{l_2} \to \bigoplus_{l=|l_1-l_2|}^{l_1+l_2}\wt{R}_l,\]
so its components, defined by the formula
\[e_{l_1,m_1}\otimes e_{l_2,m_2}=\sum (B^{-1})^{l,m}_{l_1,m_1;l_2,m_2} \wt{e}_{l,m},\label{def of A coeff}\]
are given by
\[(B^{-1})^{l,m}_{l_1,m_1;l_2,m_2}=B_{l,m}^{l_1,m_1;l_2,m_2}. \label{A in terms of B}\]
Thus the inverse transformation of the components of vectors (which is the one we actually need in the network) reads
\[\wt{v}^{l,m}=\sum B_{l,m}^{l_1,m_1;l_2,m_2}v^{l_1,m_1;l_2,m_2}.\]
This replaces the incorrect formula obtained in \citep[I.\S10.4 (p.~152)]{GelMinSha63}, which also propagated into their derivation for the Lorentz group. Now, following the derivation in \citep[II.\S6.2]{GelMinSha63} with the help of the corrected formula (\ref{A in terms of B}), we find the formula presented in the body of this paper:
\begin{multline}
    H_{(k,n),l,m}^{(k_1,n_1),l_1,m_1;(k_2,n_2),l_2,m_2}=
    \sum_{m_1',m_2'}\\ B^{\frac k2,m_1'+m_2';\frac n2,m-m_1'-m_2'}_{l,m}
    B^{\frac{k_1}{2},m_1';\frac{k_2}{2},m_2'}_{\frac{k}{2},m_1'+m_2'}
    B^{\frac{n_1}2,m_1-m_1';\frac{n_2}2,m_2-m_2'}_{\frac n2,m-m_1'-m_2'}\times \\
    \times B^{\frac{k_1}2,m_1';\frac{n_1}2,m_1-m_1'}_{l_1,m_1}
    B^{\frac{k_2}2,m_2';\frac{n_2}2,m_2-m_2'}_{l_2,m_2}.\nonumber
\end{multline}

For computational purposes, it is convenient to store an element $v^{(k,n)}$ of an irrep $T^{(k,n)}$ as a single column-vector with the combined index $M=(l,m)$ where $l=\frac{|k-n|}{2},\ldots,\frac{k+n}2$ with indices sorted over $l$ first and over $m$ last. We can thus work with vectors
\[v\in T^{(k,n)}\quad \text{ has components }v^M, \;  M=1,\ldots,(k+1)(n+1).\] 

Similarly, the CG matrix corresponding to the $(k,n)$ sector of the $(k_1,n_1)\otimes(k_2,n_2)$ product is a rectangular matrix of size $(k_1+1)(n_1+1)(k_2+1)(n_2+1)\times(k+1)(n+1)$ which can be stored as a rank 3 tensor of size $(k_1+1)(n_1+1)\times(k_2+1)(n_2+1)\times(k+1)(n+1)$:
\begin{gather}
    \prescript{(k_1,n_1),(k_2,n_2)}{(k,n)}H:\quad T^{(k,n)}\to T^{(k_1,n_1)}\otimes T^{(k_2,n_2)} \nonumber \\
    \text{Components of } H:\quad  \left(\prescript{(k_1,n_1),(k_2,n_2)}{(k,n)}{H}\right)^{M_1,M_2}_{M}.\nonumber
\end{gather}

\section{Lorentz D-matrices}

As was described in the paper, the irreps of $\SL(2,\bbC)$ can be constructed as tensor products of pairs of irreps of $\SU(2)$, that is, of pairs of Wigner-D matrices:
\[D^{k/2}(\alpha,\beta,\gamma)\otimes \overline{D^{n/2}(-\alpha,\beta,-\gamma)}.\]
However, as written these matrices act on the space  $T^{(k,0)}\otimes T^{(0,n)}$ and not $T^{(k,n)}$. Since we actually want to represent these matrices in the canonical basis of the $T^{(k,n)}$ irrep, we need to conjugate the tensor product with a matrix of CG coefficients:
\begin{multline}
    D_{(k,n)}(\alpha,\beta,\gamma)=\left(\prescript{(k,0),(0,n)}{(k,n)}{H}\right)^T\cdot \\
    \cdot \left(D^{k/2}(\alpha,\beta,\gamma)\otimes \overline{D^{n/2}(-\alpha,\beta,-\gamma)}\right)\cdot\\
    \cdot \left(\prescript{(k,0),(0,n)}{(k,n)}{H}\right).\nonumber
\end{multline}
We are not aware of a conventional name for these matrices, so for lack of a better term we call them the \emph{Lorentz D-matrices}. On $T^{(1,1)}\cong \bbR^4$, these are the familiar $4\times 4$ Lorentz matrices, i.e. the \emph{standard representation} of $\SO^+(1,3)$.

In our network, these matrices are used only to test Lorentz equivariance, but they can also be key elements of other similar architectures.

\section{Orthogonality}

Wigner D-matrices are unitary, but Lorentz D-matrices are neither unitary nor orthogonal (in fact it is known that the Lorentz group doesn't have any unitary finite-dimensional irreps). Therefore it is instructive to find a Lorentz-invariant bilinear form on all irreps. Clearly on $T^{(1,1)}$ it is the Minkowski dot product, and on other integer-spin irreps it can be induced from $T^{(1,1)}$ via tensor powers. However, invariant forms actually exist on all irreps of $\SL(2,\bbC)$. There is a prototype of a Lorentzian metric on the 2-dimensional space $R_{1/2}\cong\bbC^2$ of spinors:
\[g_{1/2}=
\begin{pmatrix}0 & 1\\
-1 & 0
\end{pmatrix}.\]
It is not Hermitian because we will be using it as a bilinear form and not as a sesquilinear form, that is, no complex conjugation is used in its definition:
\[\langle \psi,\psi'\rangle=\psi_+\psi'_--\psi_-\psi'_+.\]
Here, $\psi = (\psi_+,\psi_-)\in\bbC^2$. Thus the form can be equally viewed either as an exterior 2-form $\omega_{1/2}$ or as a pseudo-Hermitian metric $i\omega_{1/2}(\wb{\psi},\psi')$.
This form naturally induces invariant forms/metrics on all higher spin irreps of $\SU(2)$ (these forms are symmetric on integer-spin irreps). The isomorphism of representations
\[R_l\cong \left(R_{1/2}^{\otimes 2l}\right)_{\mathrm{sym}}\]
induces the forms
\[g_l=\left(g_{1/2}^{\otimes 2l}\right)_{\mathrm{sym}},\]
where the symmetrization is done separately over the two $2l$-tuples of indices. It is easy to see that in the canonical basis
\[\left(g_l\right)_{m,m'}=(-1)^{l+m}\delta_{m+m',0}.\]
For example, $g_1$ is exactly the negative of the standard Euclidean metric on $\bbR^3$.

Similarly, on the 2-dimensional irreps $(1,0)$ and $(0,1)$ of $\SL(2,\bbC)$, we choose the same form $g_{(1,0)}\!=\!g_{(0,1)}\!\coloneqq\! g_{1/2}$. Now the tensor product decomposition $T^{(k,n)}\!\cong\! \left(T^{(1,0)}\right)^{\otimes k}\otimes \left(T^{(0,1)}\right)^{\otimes n}$ induces the form
\[g_{(k,n)}=\left(g_{1/2}\right)^{\otimes k}_{\mathrm{sym}}\otimes \left(g_{1/2}\right)^{\otimes n}_{\mathrm{sym}}.\]
Another CG map can be applied to represent this product in the canonical basis, and the result is exactly the same as for $\SU(2)$ on each fixed-$l$ subspace:
\[\left(g_{(k,n)}\right)_{(l,m),(l',m')}=(-1)^{l+m}\delta_{l,l'}\delta_{m+m',0}.\]
For instance, $g_{(1,1)}$ is precisely the standard Lorentzian metric on $\bbR^4$.

CG products and D-matrices respect these forms in the sense that tensor products of two such forms generate the same forms, and D-matrices are orthogonal with respect to them (here we write this out for $\SL(2,\bbC)$ since $\SU(2)$ can be considered a special case by setting $n=0$):
\begin{gather}
    g_{(k_1,n_1)}\otimes g_{(k_2,n_2)}=\bigoplus_{(k,n)}g_{(k,n)}, \nonumber\\
    D_{(k,n)}^T g_{(k,n)} D_{(k,n)}=g_{(k,n)}.\nonumber
\end{gather}
Note that we use transposition instead of Hermitian conjugation because we treat the metric as $\bbC$-bilinear.

\section{Equivariant Universal Approximation} \label{Sec. Universality theorem}

This section provides more details on the derivation of the equivariant universal approximation theorem stated in the body of the paper.

Recall that a polynomial $f:V\to \bbR$ is called a polynomial $G$-invariant if $f(g\cdot v)=f(v)$ for all $g\in G, v\in V$. Similarly, a map $\wt{f}:V\to U$ between two representations is called a polynomial equivariant if it is equivariant and $l\circ \wt{f}$ is a polynomial for any linear functional $l:U\to \bbR$. Hilbert's finiteness theorem \cite{Hilbert1890,Hilbert1893} states that for completely reducible representations $V$ and $U$, the ring of polynomial invariants $f:V\to \bbR$ is finitely generated by a set $\{f_1,\ldots,f_{N_{\mathrm{inv}}}\}$. Similarly, all polynomial equivariants $\wt{f}: V\to U$ constitute a finitely generated module over the ring of invariants by a basis set $\{\wt{f}_1,\ldots,\wt{f}_{N_{\mathrm{eq}}}\}$ \cite{Worfolk94}. By an extension of a standard universal approximation theorem, it was shown in \cite{Yarotsky18} that for completely reducible representations, any continuous equivariant map can be approximated by a single-layer perceptron with a non-polynomial activation function $\sigma$, with the invariant generators as inputs and the equivariant generators as coefficients of the outputs. That is, there is a complete system consisting of the functions
\[
\wt{f}_i(v)\cdot \sigma\left(\sum_{j=1}^{N_{\mathrm{inv}}} w_{ij}f_j(v)+b_{i}\right),\quad i=1,\ldots,N_{\mathrm{eq}},
\]
where each of the weights $w_{ij},b_i$ spans the real line.

Therefore our network, aside from including traditional nonlinear layers acting on polynomial invariants, has to generate the basis of polynomial invariants $\{f_i\}$ and equivariants $\{\wt{f}_j\}$. 

To talk about neural networks, we adopt the definition of feed-forward neural networks from \cite{KondoTrive18}:
\begin{defn}\label{def1}
    Let $J_0,\ldots, J_L$ be a sequence of index sets, $V_0,\ldots,V_L$ vector spaces, $\phi_0,\ldots,\phi_L$ linear maps $\phi_k: V_{k-1}^{J_{k-1}}\to V_{k}^{J_{k}}$, and $\sigma_k: V_k\to V_k$ appropriate potentially nonlinear functions (acting pointwise in the sense that they are independent of the index in $J_k$). The corresponding \textit{multilayer feed-forward neural network} is then a sequence of maps $f_0, f_1, \ldots, f_L$, where $f_k=\sigma_k\circ \phi_k \circ f_{k-1}$.
\end{defn}

Now we define an equivariant analog of a feed-forward neural network.
\begin{defn}\label{def2}
    Let $G$ be a group. Let $V_0,\ldots,V_{2L}$ be finite-dimensional vector spaces that are also linear representations of $G$, $\sigma_k:V_k\to V_{k+1}$, $k=0,2,\ldots,2(L-1)$, -- potentially nonlinear $G$-equivariant maps, and $\phi_k:V_k\to V_{k+1}$, $k=1,3,\ldots,2L-1$, -- $G$-equivariant linear maps. Then the corresponding \textit{$G$-equivariant multilayer feed-forward neural network} is the sequence of maps $f_0,\ldots,f_L$, where $f_k=\phi_{2k+1} \circ \sigma_{2k}\circ f_{k-1}$.
\end{defn}

\begin{defn}
    A \textit{polynomial $G$-equivariant feed-forward neural network} is a $G$-equivariant one in the sense of Def.~\ref{def2} in which all nonlinearities $\sigma_k$ are polynomial. Specifically, all such $\sigma_k$ can be expressed using tensor products and $G$-equivariant linear maps. A minimal example with a quadratic nonlinearity is $\sigma_k(v)=v\oplus (v\otimes v)$.
\end{defn}

\begin{lem}\label{lem polynomials}
    If $\sigma:V\to U$ is a polynomial $G$-equivariant map of degree $d$ between two completely reducible finite-dimensional representations $V,U$ of $G$, then there exist $G$-equivariant maps $\alpha_p:V^{\otimes p}
    \to U$, $p=0,\ldots,d$, such that 
    \begin{equation}
        \sigma = \sum_{p=0}^d \alpha_p \left(v^{\otimes p}\right).\label{polynomial rep}
    \end{equation}
\end{lem}
\begin{proof}
    Decompose $\sigma$ into homogeneous components $\sigma=\sum_{i=0}^d p_i$. Since the action of $G$ is linear, each $p_i$ separately is $G$-equivariant: $p_i(\rho_V(g)\cdot v)=\rho_U(g)\cdot p_i(v)$. Thus, without loss of generality, we can assume that $\sigma$ is homogeneous.
    
    If $\sigma$ is homogeneous of degree $d$, it can be written as 
    \[\sigma(v)=p(\underbrace{v,\ldots,v}_d)\]
    for some symmetric $d$-multilinear map $p:V^{d}\to U$. Such a multilinear map is identified with an element of the tensor product space
    \begin{equation}
        t\in S^d(V^\ast) \otimes U, \label{tensor space}
    \end{equation}
    where $S^d(V^\ast)=\left(V^\ast\right)^{\otimes d}_{\mathrm{Sym}}$ is the symmetric tensor power of $V^\ast$. Therefore all polynomial equivariants on $V$ are indeed tensor polynomials, i.e. $p$ can be viewed as a \textit{linear} equivariant map $p:V^{\otimes d}\to U$. Since this tensor is symmetric, this proves the existence of a linear equivariant $\alpha_d$ such that $\sigma(v)=\alpha_d\left(v^{\otimes d}\right)$.
\end{proof}

\begin{lem}
    Given two completely reducible finite-dimensional representations of a group $G$, the space of polynomial $G$-equivariant maps from $V$ to $U$ is isomorphic to the subspace of invariants in the tensor product $S(V^\ast) \otimes U$, where $S(V^\ast)$ is the symmetric tensor algebra over $V^\ast$:
    \[
    \mathrm{Pol}_G(V,U)\cong \left(\left(S(V^\ast\right)\otimes U\right)^G.
    \]
\end{lem}
\begin{proof}
    As shown in the proof of Lemma \ref{lem polynomials}, there is an isomorphism with the space of $G$-equivariant linear maps mapping $S(V^\ast)\to U$:
    \[\mathrm{Pol}_G(V,U)\cong \Hom_G (S(V),U).\]
    Since the hom-functor is the adjoint of the tensor product functor, we have 
    \[\Hom_G (S(V),U)\cong \Hom_G (S(V)\otimes U^\ast,\bbR)=(S(V^\ast)\otimes U)^G.\]
    See also \cite{Miller71}.
\end{proof}

\begin{rem}\label{remark1}
    The computation of this space clearly comes down to finding an isotypic decomposition of the tensor algebra over $V$ (we expand on this in Remark \ref{remark irr}). The isotypic decomposition of the symmetric tensor algebra $S(V^\ast)$ thus provides a complete system of polynomial equivariants. Namely, assuming without loss of generality that $U$ is irreducible, any $\sigma\in \mathrm{Pol}_G(V,U)$ can be written as in (\ref{polynomial rep}), where each $\alpha_p$ is a composition $\alpha_p=\beta_p\circ P_U^p$ of the projector $P_U^p:V^{\otimes p}\to U^\tau$ onto the $U$-type isotypic component of $V^{\otimes p}$ and a $G$-equivariant linear map $\beta_p:U^\tau\to U$.
\end{rem}

These lemmas imply that the seemingly nonlinear problem of constructing all polynomial equivariants on $V$ can be reduced to the \textit{linear} problem of computing the isotypic decompositions of tensor powers of $V$. We now state more precisely our equivariant approximation theorem.

\begin{thm}
    Let $G$ be a classical Lie group and $V,U$ two completely reducible finite-dimensional representations of $G$. Then any continuous equivariant map $F:V \to U$ can be uniformly approximated by equivariant feed-forward neural networks in the sense of Def.~\ref{def2}, in which all nonlinearities are based on tensor products, except perhaps when acting on $G$-invariants. For example, given a non-polynomial function $\wt{\sigma}_k:\bbR\to \bbR$, we can have 
    \begin{equation}
        \sigma_k(v)=\wt{\sigma}_k\left(P_{\mathrm{inv}}(v)\right)\oplus v\oplus(v\otimes v),\label{nonlinearity}
    \end{equation}
    where $P_{\mathrm{inv}}$ is the projector onto invariants and the action of $\wt{\sigma}_k$ on the vector of invariants is component-wise.
\end{thm}
\begin{proof}

This theorem follows immediately from Remark \ref{remark1}. Indeed, \citet{Yarotsky18} showed that, given a basis of polynomial invariants and equivariants, a conventional neural network can uniformly approximate an equivariant function. We have further demonstrated that such a basis can be generated up to an arbitrary polynomial degree by an equivariant feed-forward neural network which can construct all possible tensors of the inputs and compute the isotypic components of these tensors. A nonlinearity such as (\ref{nonlinearity}) iterated sufficiently many times constructs a basis for all tensors of $v$ and applies scalar nonlinearities to all $G$-invariants.
\end{proof}

\begin{rem}\label{remark irr}
Here we further specify the form of the equivariant tensors constructed above.
Since $V$ admits a decomposition into a direct sum $V \cong \bigoplus_{i} V_{\alpha_i}$ of irreps $R_{\alpha_i}$ labeled by their highest weight $\alpha_i$, then an equivariant $\wt{f}:V\to U$ viewed as a function of several vectors $f(v_1,v_2,\ldots)$ with $v_i\in V_{\alpha_i}$, has to be a homogeneous polynomial of some degree $k_i$ in each $v_i$. As shown in the Lemmas above, this allows one to view $\wt{f}$ as a \textit{multilinear} $U$-valued function $t$ of $\sum_i k_i$ vectors, where each $v_i$ is repeated $k_i$ times:
\[
\wt{f}(v_1,v_2,\ldots)=t(\underbrace{v_1,\ldots,v_1}_{k_1\text{ times}}\underbrace{v_2,\ldots,v_2}_{k_2\text{ times}},\ldots).
\]
Just like in the proof of Lemma \ref{lem polynomials}, this multilinear function can be interpreted as an element of the symmetric tensor product
\begin{equation}
t\in \left(\bigotimes_i \left(R_{\alpha_i}^\ast\right)^{\otimes k_i}_{\mathrm{Sym}}\right) \otimes U. \label{tensors}
\end{equation}

Assuming without loss of generality that $U=R_\alpha$ is an irrep, the problem of constructing all equivariants $V\to R_\alpha$ is reduced to computing the $R_\alpha$-isotypic component of this tensor algebra.
\end{rem}

More information on these constructions in classical invariant theory can be found in e.g.~\cite{GoodmWalla09} and \cite{Weyl46}.
As a side note, we restate the following classical theorem \citep[Thm.~5.5.21]{GoodmWalla09}:
\begin{thm*}
    If $G$ is a classical Lie group, say, $\SU(2)$, $\SL(2,\bbC)$, $\SO(3)$, or $\SO^+(1,3)$, and $V$ is its fundamental representation (of dimension 2, 2, 3, and 4, respectively), then any finite-dimensional irrep of $G$ occurs as a $G$-invariant subspace of the tensor power $V^{\otimes k}$ for a sufficiently high $k$.
\end{thm*}

Considering the case of the Lorentz group, taking all possible tensor products of input 4-vectors and decomposing into irreducibles we will generate tensors that transform under arbitrary irreps of the group. Therefore there are no restrictions on the type of equivariant outputs that our architecture can produce. In fact, the dimensions of the spaces of equivariants mapping a set of 4-vectors to an irrep $U=R_\alpha$ of the Lorentz group are known \cite{Miller71}.

\section{Equivariance Tests}

We have conducted experiments to verify Lorentz invariance of our neural network. The network itself had exactly the same values of hyper-parameters as in the main application, but the inputs were replaced by random 4-momenta, whose components are drawn uniformly from $[-1,1]$, with 20 particles in each event and 20 events in a batch. The outputs of the network are then arrays $w$ of shape $2\times 20$. We compute the outputs for the same 4-momentum inputs with and without a Lorentz matrix applied to them at the start. Calling these two outputs $w$ and $\wt{w}$, we define the relative deviation as $\mathrm{mean}(w-\wt{w})/\mathrm{mean}(w)$. We computed these quantities for a number of Lorentz boosts with varying Lorentz factor $\gamma$ and averaged the results over 10 sets of random inputs and random initializations of the model (60 events with 20 particles each in total). The computations here are done using double precision and the relative error remains within $0.1\%$ up to gamma factors of about $5000$, which well covers the physically relevant domain of about $[10,200]$. When using 32 bit precision, the error remains this low only up to $\gamma\sim 70$ and grows to over 10\% after $\gamma\sim 200$.

\begin{figure}[ht]
\begin{center}
\centerline{\includegraphics[width=\columnwidth]{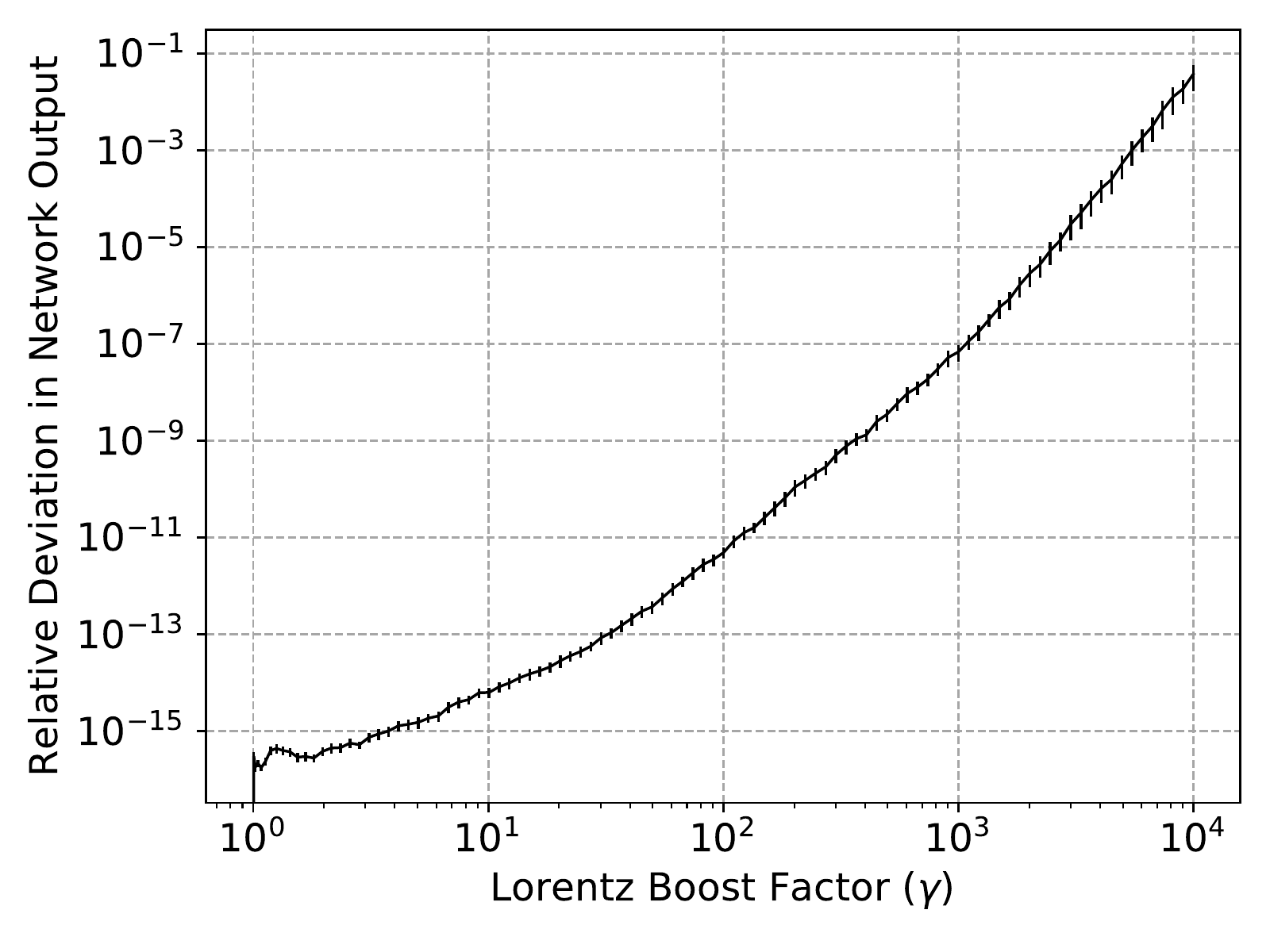}}
\caption{Relative deviation of the outputs of the network as a function of the boost factor $\gamma$ applied to its inputs.}
\end{center}
\vskip -0.2in
\end{figure}
 
Similarly we have tested rotational invariance, however the error is remains strictly of the order $10^{-16}$ when using double precision (the Euler angle of the rotation ranged from 0 to 10), so we are not showing a separate plot for it. It is clear that the source of the error is just the rounding errors in float arithmetic, so larger inputs produce larger relative errors. That is why applying large boosts increases the error, but rotations do not have the same effect.

Finally, the internal equivariance of the network was tested as well by applying Lorentz matrices to the inputs and comparing the values of the resulting activations of the network to an application of corresponding Lorentz D-matrices to them. The errors are similarly small, so we do not show separate statistics for them.

\section{Computational Cost}
\begin{figure}[ht]
\begin{center}
\centerline{\includegraphics[width=\columnwidth]{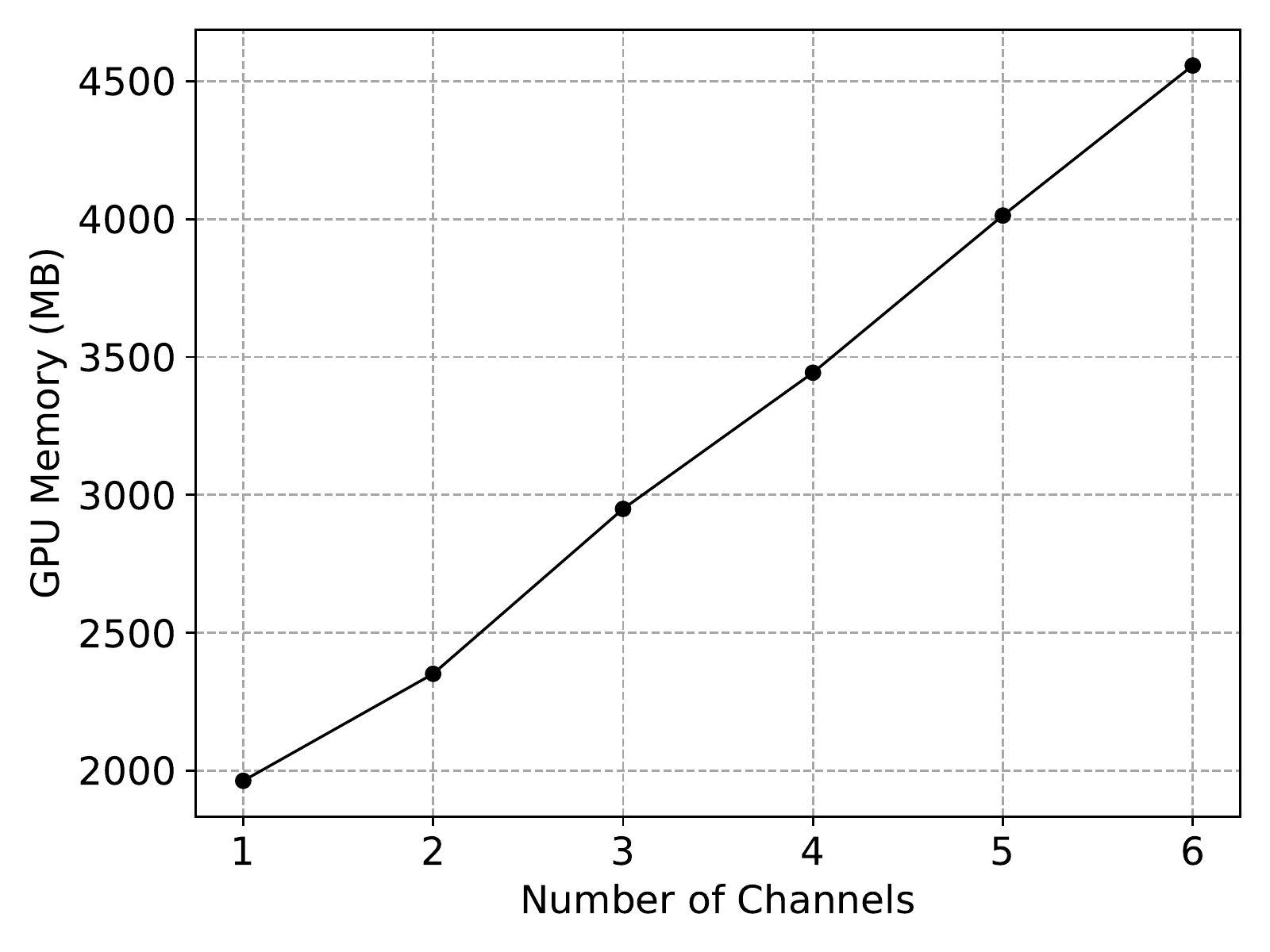}}
\caption{GPU memory usage as a function of the number of channels per layer, with 3 layers. \label{fig:gpumemory}}
\end{center}
\vskip -0.2in
\end{figure}

\begin{figure}[ht]
\begin{center}
\centerline{\includegraphics[width=\columnwidth]{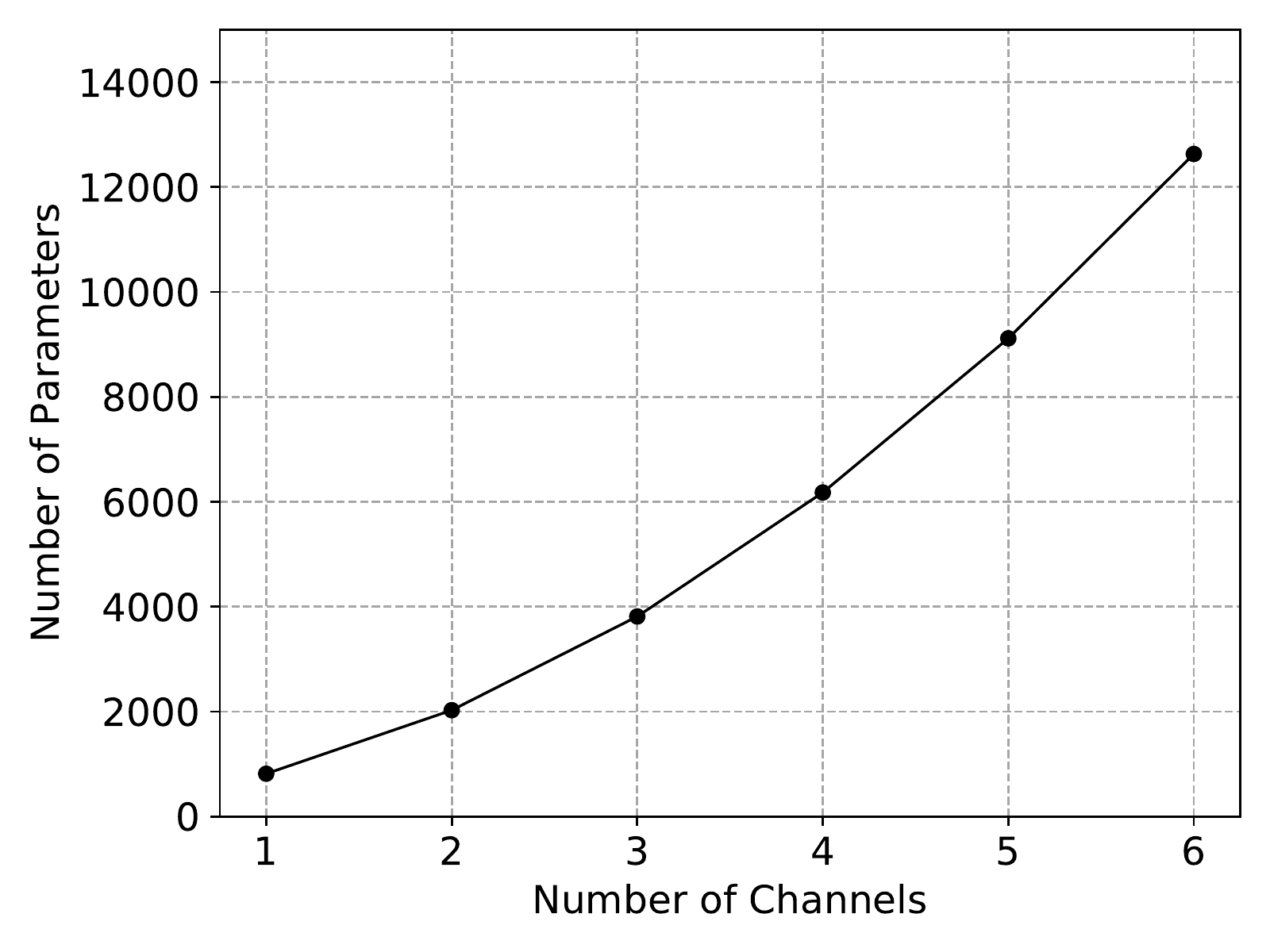}}
\caption{The number of network parameters as a function of the number of channels per layer, with 3 layers. \label{fig:parameters}}
\end{center}
\vskip -0.2in
\end{figure}

Here we present the plots of the GPU memory (Fig.~\ref{fig:gpumemory}) and the number of parameters (Fig.~\ref{fig:parameters}) as functions of the number of channels (which here is uniform across all layers). These numbers correspond to the same model as the one trained for our main experiment, except for the modified number of channels. We note that the usage of GPU memory is much more efficient when the sizes of all tensors are multiples of 32. The size of most tensors is $2\times B\times N_{\mathrm{obj}}^{s}\times N_{\mathrm{ch}}\times d$ with $B$ being the batch size, $N_{\mathrm{obj}}$ the number of particles (202 for the top-tagging dataset), the power $s=1$ or $2$, and $d$ the dimension of an irrep. The number of model parameters grows roughly quadratically with the number of channels.

Since the sizes of some of the tensors involved grow quadratically with the number of particles $N_{\mathrm{obj}}$, and we take tensor products of them, the evaluations of this model take a much longer time than simpler models. This can be mitigated by optimizing the tensor product operation. Namely, since Clebsch-Gordan coefficients satisfy several symmetry relations and ``conservation laws'', one may replace the tensor product followed by the CG operation with a single operation performed efficiently on the GPU. A custom CUDA kernel for this purpose is under development.

\section{Network Metrics}

\begin{figure}[h]
\begin{center}
\centerline{\includegraphics[width=\columnwidth]{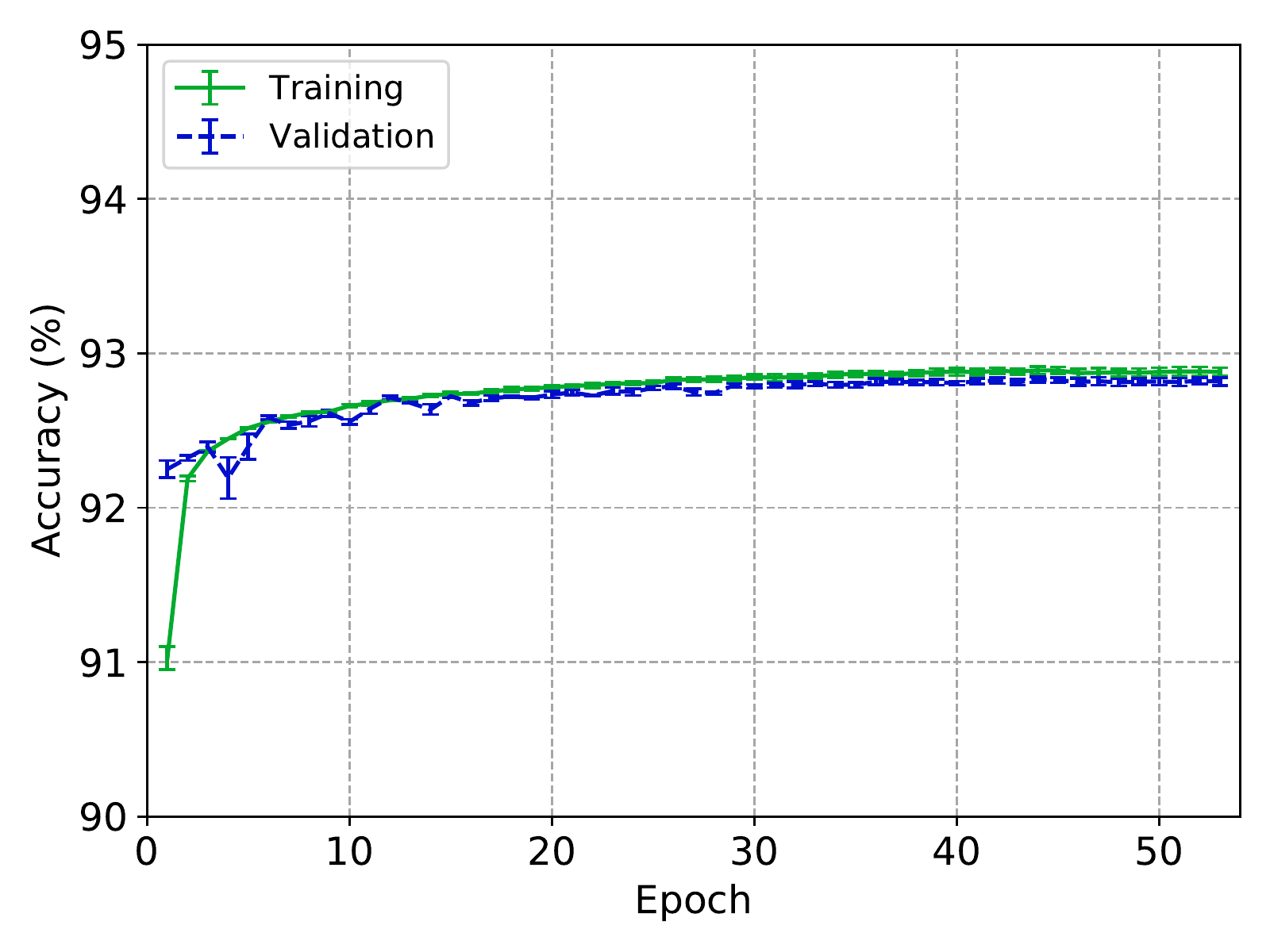}}
\caption{The average network accuracy as a function of epoch number, sampled over 4 independent trained instances. The two data series correspond with results from the training and validation subsets of the dataset \cite{KasPleThRu19}. The error bar width is given by the standard deviation. \label{fig:accuracy}}
\end{center}
\vskip -0.2in
\end{figure}

\begin{figure}[h]
\begin{center}
\centerline{\includegraphics[width=\columnwidth]{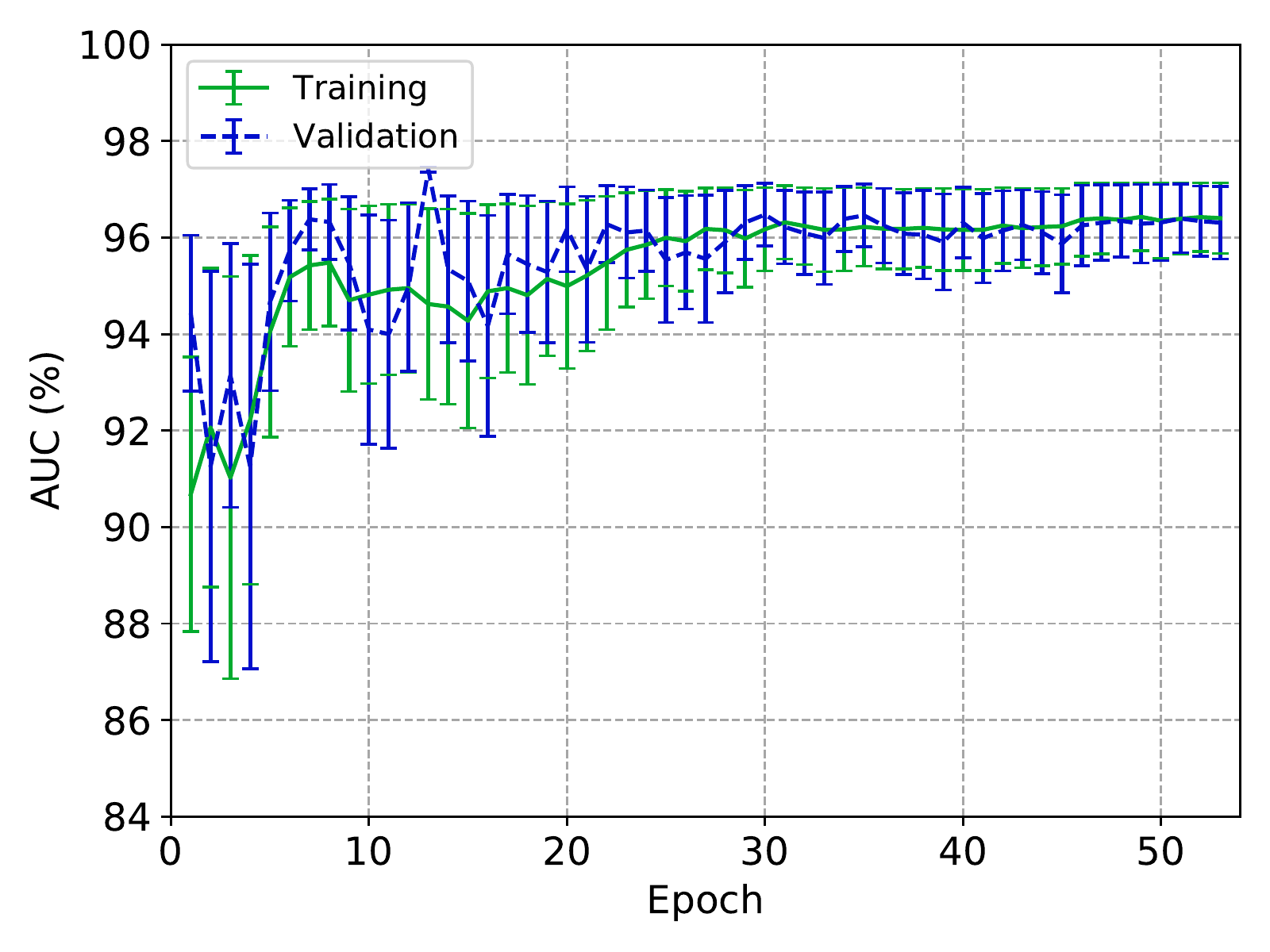}}
\caption{The average area under the ROC curve (AUC), as a function of epoch number. The error bar width is given by the standard deviation. \label{fig:auc}}
\end{center}
\vskip -0.2in
\end{figure}

\begin{figure}[h]
\begin{center}
\centerline{\includegraphics[width=\columnwidth]{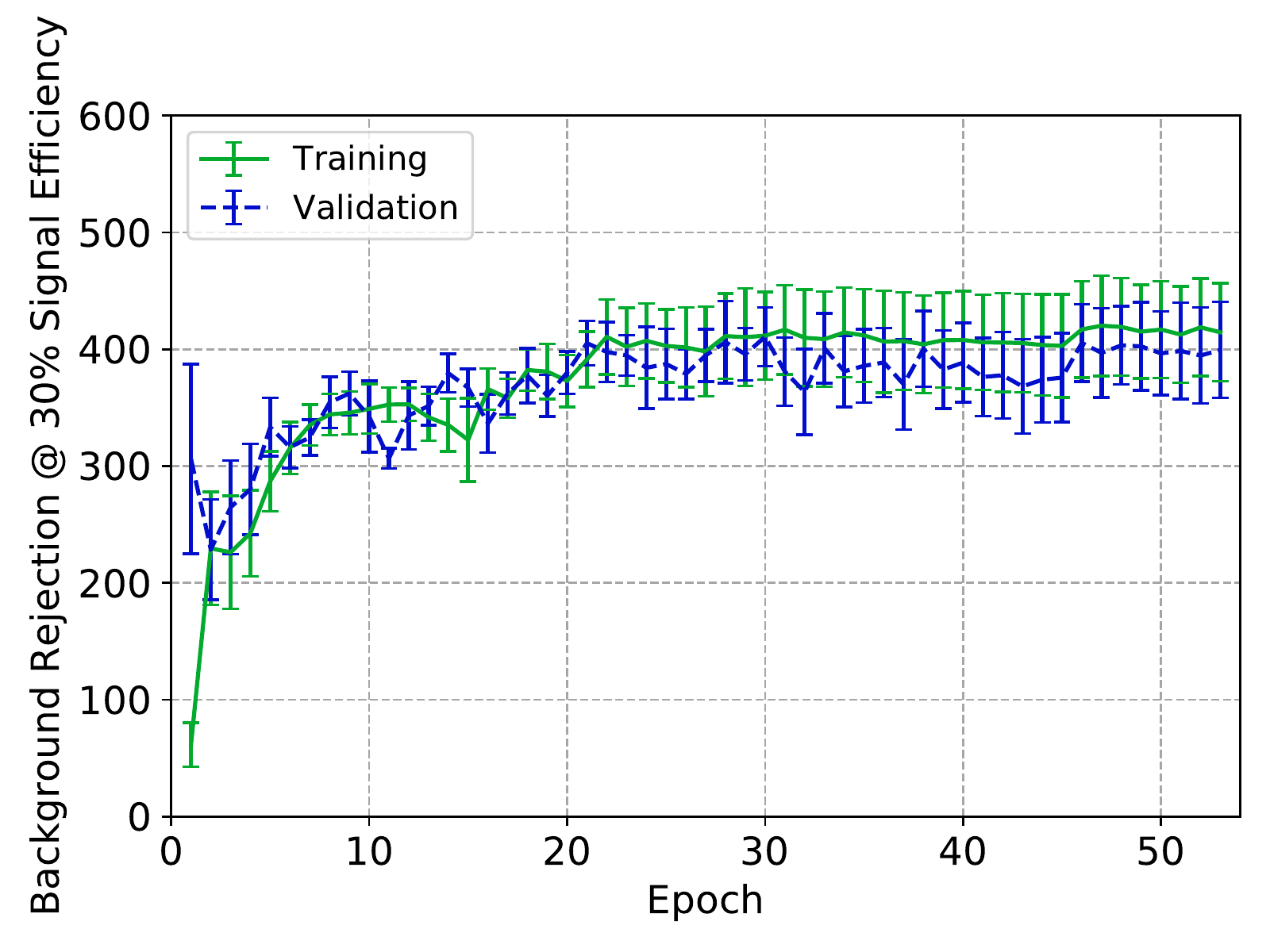}}
\caption{The average background rejection at 30\% signal efficiency, as a function of epoch number. The error bar width is given by the standard deviation. \label{fig:rejection}}
\end{center}
\vskip -0.2in
\end{figure}

\begin{figure}[h]
\begin{center}
\centerline{\includegraphics[width=\columnwidth]{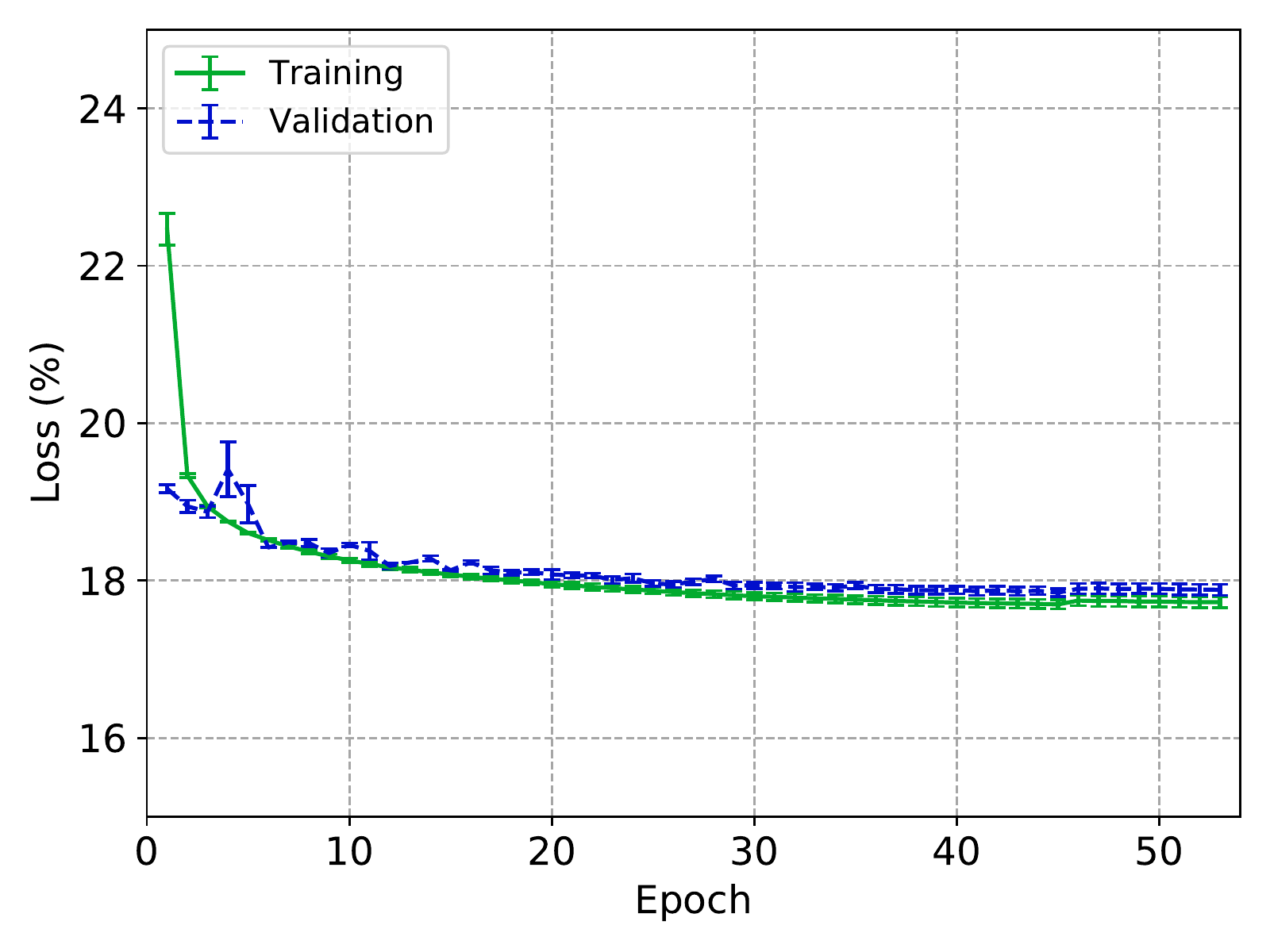}}
\caption{The average loss, as a function of epoch number. The error bar width is given by the standard deviation. \label{fig:loss}}
\end{center}
\vskip -0.2in
\end{figure}

Lastly, we display the evolution of some of the metrics of the network with the number of epochs -- these were measured from the ensemble of networks from our main experiment. The accuracy (Fig~\ref{fig:accuracy}) and AUC (Fig~\ref{fig:auc}) score appear to reach a rough ceiling partway through training, whereas the background rejection (Fig~\ref{fig:rejection}) and loss (Fig~\ref{fig:loss}) continue to improve throughout.

\section{Source Code}

The source code is available at \url{https://github.com/fizisist/LorentzGroupNetwork}. It requires PyTorch and CUDA for training on a GPU (not yet parallelized across multiple GPU's). It also uses NumPy and Scikit-Learn for some diagnostics, and H5py for reading data from HDF datasets.

\newpage

\end{document}